\documentclass[12pt]{iopart}
\usepackage{iopams,graphicx}
\begin{document}

\title[Short GRBs and neutron star mergers]{Short gamma-ray bursts as a
possible probe of binary neutron star mergers}
\author{Shin'ichiro Ando}

\address{Department of Physics, School of Science, The University of
Tokyo, 7-3-1 Hongo, Bunkyo-ku, Tokyo 113-0033, Japan}

\ead{ando@utap.phys.s.u-tokyo.ac.jp}

\begin{abstract}

We study the properties of short gamma-ray bursts (GRBs), based on the
 assumption that they are all connected to the binary neutron star
 mergers, whose formation mechanism contains a large amount of
 uncertainty.
In particular, the merger time distribution is focused on, which is
 essential for short GRB rate estimate; we simply parameterize it as
 $P_{\rm m}(t)\propto t^\alpha$ with lower cutoff time-scale $\tau$.
The short GRB rate, calculated using such a parameterization with a
 specific model of cosmic star formation rate, is found to be quite
 insensitive to the value of $\tau$, but considerably sensitive to
 $\alpha$.
With such formation rate models, the peak flux distribution observed by
 BATSE is fitted.
In the fitting process, the local short GRB rate $\rho_0$ is treated as
 one of free parameters, giving the best fit when $\rho_0 = 0.1$--2
 yr${}^{-1}$ Gpc${}^{-3}$, depending on the other free parameters as
 well as the formation rate model represented by $(\tau,\alpha)$.
We also discuss several implications for the intrinsic properties of
 possible sources of short GRBs, such as jet structure, intrinsic
 luminosity function, and intrinsic duration distribution.

\end{abstract}



\maketitle

\section{Introduction}
\label{sec:Introduction}

In recent years, our knowledge of gamma-ray bursts (GRBs) enormously
improved through many observations of GRBs themselves (using gamma-rays)
as well as their transient component, i.e., afterglow, which can be
detected in various wavebands from radio to X-rays.
Especially from the afterglow observations, their cosmological origin
has been established, which revealed that GRBs are very luminous
astrophysical objects whose isotropic energy emitted by gamma-rays
ranges $10^{52}$--$10^{54}$ erg.
Furthermore, evidence of jetted GRBs arises from long term radio
observations \cite{Waxman98} and from observations of achromatic breaks
in the afterglow light curves \cite{Rhoads97}.
Using the observed break time, the jet opening angle and intrinsic
gamma-ray energy were inferred \cite{Frail01}, and the authors have
concluded that the gamma-ray energies, after correcting the beaming
effect, are clustered around $\sim 10^{51}$ erg.
However, the remarkable progress is restricted only to the long duration
class of GRBs, whose time duration $T_{90}$ exceeds 2 s, where $T_{90}$
is the interval over which 5--95\% of the burst counts accumulate.
On the other hand, no afterglows have been observed thus far for short
duration GRBs ($T_{90}<2$ s), and as its consequence, there is no direct
knowledge of their redshifts and intrinsic luminosities.

Because of its energy and time scales, GRB has been considered to be
connected to massive compact stars, such as supernova explosions and
binary neutron star mergers.
In fact, several observations indicate that long GRBs are likely
associated with supernova explosions \cite{Galama98}; the best evidence
for the connection is the spectroscopic discovery of SN 2003dh in the
afterglow spectrum of GRB 030329 \cite{Stanek03}.
The absence of afterglow observation for the short GRBs, on the other
hand, restricts our discussions concerning their origin.
However, it is natural to interpret that the phenomenon that triggers
short GRBs is different from that triggering long GRBs; this is due to
clear bimodality of the distribution of $T_{90}$ accumulated in the
BATSE data \cite{Kouveliotou93}.
In the present paper, we assume that the short GRBs are associated with
the binary neutron star mergers, i.e., neutron star--neutron star
(NS-NS) or neutron star--black hole (NS-BH) coalescence.
Numerical studies actually suggest that short duration burst can be
explained by the NS-NS or NS-BH merger models \cite{Popham99}.

Coalescence rates of binary compact object system have been estimated in
two ways: (1) theoretically, based on the predictions of binary
population synthesis calculations (e.g.,
\cite{Lipunov95,Fryer99b,Belczynski02b})
or (2) empirically, based on the observed sample of galactic binary
pulsars (e.g., \cite{Narayan91}).
At present, both methods appear burdened with significant uncertainties:
the theoretical approach due to the many poorly constrained evolutionary
model parameters and the empirical estimates due to small number sample
of observed NS-NS systems.
Thus, both theoretically and observationally estimated coalescence rates
considerably change as a result of discovery of (theoretical) new
merger formation channels or (observational) tight binary systems; in
fact, there were such discoveries in both approaches.
In the theoretical population synthetic approach, a new formation
channel was found to be relevant in reference \cite{Belczynski02a},
resulting in the increase of the number of tight binary compact object
systems that possibly merge in less than 1 Myr from their formation.
From the observational aspect, a highly relativistic binary neutron star
system has been recently discovered \cite{Burgay03}, which will merge in
about 85 Myr as a result of gravitational wave emission, and the
observational rate estimate increased by a factor of 6--7 compared with
the previous one \cite{Kalogera03}.

As we assume in the present study, if the short GRBs are truly
associated with binary neutron star mergers, it is expected that the
observations of short GRBs and their afterglows would provide rich
information on the picture of binary systems, such as formation rate
history and surrounding environment.
In this paper, we investigate the formation rate history of short GRBs
using the peak flux distribution obtained by the BATSE detector,
particularly focusing on the merging time distribution (for an earlier
approach, see reference \cite{Piran92}).
It depends on the physical processes of the binary formation that are
still quite uncertain as already discussed in the previous paragraph.
Several past studies \cite{Totani97b,Wijers98,Porciani01} have also used
the peak flux distribution, but in order to obtain implications for
cosmic star formation history or luminosity functions (LFs) of
long/short GRBs.
In the present paper, we rather aim at deriving physical properties of
NS-NS or NS-BH mergers from the short GRB observation, assuming the
connections between them and that the observationally inferred star
formation rate (SFR) correctly represents nature.
Although there remain huge amount of uncertainties concerning the cosmic
SFR history, many observations with various wavebands seem to be roughly
consistent with each other at low-redshift region after some correction
law of dust extinction is applied.

Because the available data are quite restricted, owing to absence of
any afterglow signals, we believe that it would be better to use simple
parameterization for merger time distribution as described in
\sref{sec:Formation history of short gamma-ray bursts}.
It is found that the GRB rate history is quite sensitive to the merger
time distribution.
In \sref{sec:Peak flux distribution and theoretical models}, we briefly
model necessary properties of short GRBs such as the LF and photon
spectrum.
The available BATSE data as well as its efficiency are also summarized
in the same section.
\Sref{sec:Results of parameter fitting} is devoted to showing the
results of parameter fitting.
The best fit values for each parameter as well as its allowed region at
$1\sigma$ level are presented.
Finally in \sref{sec:Discussion}, we discuss the obtained properties
concerning neutron star mergers as a potential source of short GRBs, by
comparing our result with those obtained by other approaches such as
population synthesis; we discuss the intrinsic GRB properties such as
lumnosity and jet structure.
In addition, the redshift distribution of BATSE short bursts is inferred
using the best fit parameters, and then we derive the intrinsic duration
distribution by correcting time dilation effect.

\section{Formation history of short gamma-ray bursts}
\label{sec:Formation history of short gamma-ray bursts}

\begin{figure}[htbp]
\begin{center}
\includegraphics[width=10cm]{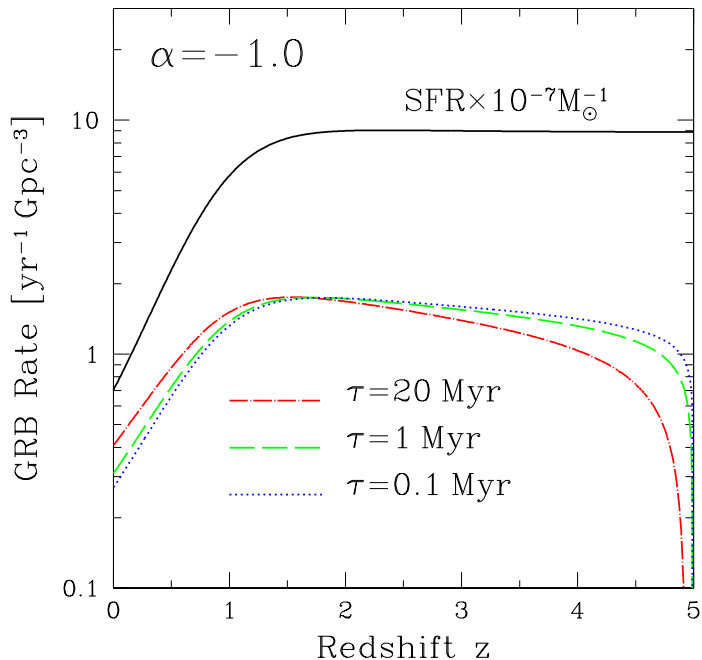}
\caption{Formation rate history of short GRBs. Each rate history is
 obtained with the assumption that the short GRBs are associated with
 binary neutron star mergers. Dot-dashed, dashed and dotted curves
 correspond to different lower cutoff time-scale of coalescence
 distribution $P_{\rm m}(t)$ with its slope fixed to be $\alpha =-1.0$.
 The absolute values are normalized to best fit the observed peak flux
 distribution as discussed in \sref{sec:Results of parameter fitting}.
 Cosmic SFR density, from which GRB rates are generated, is also shown
 as solid curve. \label{fig:GRBrate1}}
\end{center}
\end{figure}

By recent progressive observations at various wavebands, the cosmic SFR
is fairly well known, although there remains debate on especially
high-redshift universe.
As our reference model, we use a simple functional form for the SFR per
comoving volume, which is given in reference \cite{Porciani01} (labeled
as `SF2'), as
\begin{equation}
 \fl R_{\rm SF2} (z)=0.16h_{70}\frac{\exp (3.4z)}{\exp (3.4z)+22}
  \frac{[{\Omega_{\rm m}(1+z)^3+\Omega_\Lambda}]^{1/2}}{(1+z)^{3/2}}
  \,M_\odot\,{\rm yr}^{-1}\,{\rm Mpc}^{-3}\, ,
\end{equation}
and we adopt the standard $\Lambda$CDM cosmology ($\Omega_{\rm m}=0.3,
\Omega_\Lambda=0.7,H_0=70\,h_{70}$ km s${}^{-1}$ Mpc${}^{-1}$);
\fref{fig:GRBrate1} shows SFR-$z$ relation with this model (solid
curve).
The SFR sharply increases from the current epoch $z=0$ to $z\simeq 1.5$,
and then it keeps the same level to farther past universe, $z=5$.
Behaviours at high-redshift region are quite uncertain because the
galaxy LF has not been established yet there as well as the amount of
dust extinction is highly ambiguous.
Therefore, it may be useful to compare several SFR models which differ
at high redshift from each other, as actually done in reference
\cite{Porciani01} (for long GRBs).
In the present study, we further adopt two other SFR models also given
in reference \cite{Porciani01} as `SF1' and `SF3', and compare the
results obtained by these models to that by the standard model SF2.
The models SF1 and SF3 are given as a functional form by
\begin{equation}
\fl R_{\rm SF1} (z)=0.32h_{70}\frac{\exp (3.4z)}{\exp (3.8z)+45}
  \frac{[{\Omega_{\rm m}(1+z)^3+\Omega_\Lambda}]^{1/2}}{(1+z)^{3/2}}
  \,M_\odot\,{\rm yr}^{-1}\,{\rm Mpc}^{-3}\, ,
  \label{eq:SF1}
\end{equation}
\begin{equation}
\fl R_{\rm SF3} (z)=0.22h_{70}\frac{\exp (3.05z-0.4)}{\exp (2.93z)+15}
  \frac{[{\Omega_{\rm m}(1+z)^3+\Omega_\Lambda}]^{1/2}}{(1+z)^{3/2}}
  \,M_\odot\,{\rm yr}^{-1}\,{\rm Mpc}^{-3}\, .\\
 \label{eq:SF3}
\end{equation}
Behaviours of these two functions are roughly the same as that of SF2 at
low-redshift region $z<1.5$, but on the other hand, they significantly
differ at $z>1.5$; at that region, the SF1 and SF3 models give
exponentially decreasing and increasing formation rate, respectively,
compared with the constant behaviour of SF2.
In fact, as shown in \sref{sec:Results of parameter fitting}, it is
found that the results obtained by the SF1 and SF3 models are
essentially unchanged compared with that by SF2, since the bulk data
should be contributed rather low redshift region $z\lesssim 1.5$, at
which three SFR models are only slightly different.

The short GRB rate, $R_{\rm GRB}(t)$, can be calculated assuming its
association with binary neutron star mergers, by using the formula
\begin{equation}
 R_{\rm GRB}(t)\propto\int_{t_{\rm F}}^t
  \rmd t^\prime R_{\rm SF} (t^\prime)P_{\rm m}(t-t^\prime)\, ,
  \label{eq:short GRB rate}
\end{equation}
where $t_{\rm F}$ represents the formation epoch of galaxies, about
which we assume $z(t_{\rm F})=5$.
$P_{\rm m}(t)$ is the probability distribution function of merging time
of the binary system from its formation.
This distribution can be calculated theoretically using population
synthesis method \cite{Lipunov95,Fryer99b,Belczynski02b}, but resulted
in various different distributions depending on studies, which means
that quite large ambiguities should be accompanied.
In the present paper, therefore, we take more simplified approach; i.e.,
a simple parameterization such as $P_{\rm m}(t)\propto t^\alpha$ with
lower cutoff time-scale $\tau$ is adopted.
This is because we believe that this approach would be more useful to
obtain some intuitive implications.
We adopt as the power law index $\alpha$ values ($-0.5,-0.7,-1.0,-1.5$),
and ($0.1,1,20$) as $\tau_{\rm Myr}$, where $\tau_{\rm Myr}$ represents
the lower cutoff time-scale in unit of Myr.
One of the parameter sets we adopt, i.e., $(\alpha,\tau_{\rm
Myr})=(-1.0,20)$, was also investigated in reference \cite{Totani97b}.
As already mentioned in \sref{sec:Introduction}, a new formation channel
was pointed out in reference \cite{Belczynski02a}, with which a large
amount of tight binaries, which merge within 1 Myr from their formation,
possibly form.
It resulted in decrease of lower cutoff time-scale $\tau$; for instance,
merging time distribution calculated in reference \cite{Belczynski02b}
is found to peak at $\sim 0.3$ Myr, which is significantly shorter than
the previously considered value, i.e., $\gtrsim$ 10--100 Myr.

\begin{figure}[htbp]
\begin{center}
\includegraphics[width=10cm]{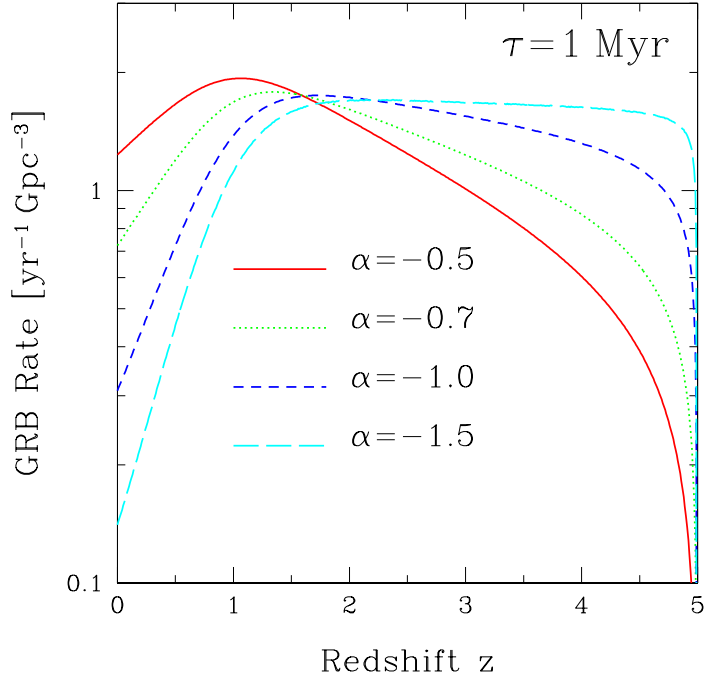}
\caption{The same as \fref{fig:GRBrate1}, but evaluated for various
 values of $\alpha$ with fixed $\tau$ to be 1
 Myr.\label{fig:GRBrate2}}
\end{center}
\end{figure}

Figures \ref{fig:GRBrate1} and \ref{fig:GRBrate2} shows the short GRB
rate density as a function of redshift due to binary neutron star
mergers that is calculated with equation \eref{eq:short GRB rate}.
Normalization of the absolute value is obtained by the parameter
fitting that is discussed in \sref{sec:Results of parameter fitting}.
Dependence on the parameter $\tau$ is shown in \fref{fig:GRBrate1} with
$\alpha=-1.0$.
As the value of $\tau$ is increased, the local rate evolution to $z\sim
1$ becomes shallower than that of original SFR model, from which the
short GRB rate was generated via equation \eref{eq:short GRB rate}.
This is because the time delay effect from the formation to coalescence
is more prominent in the case of larger $\tau$.
On the other hand, in \fref{fig:GRBrate2} we compare models with
different values for power law index $\alpha$ assuming $\tau_{\rm Myr} =
1$; the GRB rate history is found to change more dramatically depending
on the value of $\alpha$.
Using smaller $\alpha$, which increases the relative contribution of
tight binaries to the total rate, the GRB rate becomes steeper at lower
redshift; especially for $\alpha\lesssim -1.5$, the GRB rate history
becomes quite good tracer of cosmic SFR.
In consequence, the short GRB rate history is quite sensitively
dependent on the unknown parameters of binary neutron star systems.

\section{Peak flux distribution and theoretical models}
\label{sec:Peak flux distribution and theoretical models}

We define short GRBs as those with $T_{90}<2$ s, and take from the BATSE
current burst catalogue.\footnote{http://cossc.gsfc.nasa.gov/batse/}
The BATSE peak flux is given in three different time-scales of 64, 256
and 1024 ms.
We use the 64 ms triggering timescale because the average duration of
short GRBs is $\sim 0.3$ s, and we have to use a shorter timescale than
this for the real peak flux.

The peak flux distribution can be theoretically calculated using the
following formulation:
\begin{equation}
 \frac{\rmd\dot N}{\rmd P}=\int_0^{\infty}
  \rmd z\frac{\rmd V(z)}{\rmd z}\frac{R_{\rm GRB}(z)}{1+z}
  \psi (L(P,z))\epsilon (P)\frac{\partial L(P,z)}{\partial P}\, ,
  \label{eq:N-P distribution}
\end{equation}
which is in unit of sr${}^{-1}$ yr${}^{-1}$ cm${}^2$ s.
Here, $\rmd V/\rmd z$ is the comoving volume element per unit solid
angle, $\psi (L)$ is the peak LF of short GRBs, $\epsilon (P)$ is the
detector efficiency as a function of photon flux, and the factor
$(1+z)^{-1}$ accounts for cosmological time dilation.
In our notation, the burst peak luminosity $L$ is given as a function of
redshift $z$ (that corresponds to distance) as well as the observed peak
flux $P$.
The comoving volume element is simply represented by the following
formula with cosmological parameters:
\begin{equation}
 \frac{\rmd V}{\rmd z}=\frac{c}{H_0(1+z)^2}
  \frac{d_{\rm L}^2(z)}{[\Omega_{\rm m}(1+z)^3+\Omega_\Lambda]^{1/2}}\,,
  \label{eq:comoving volume}
\end{equation}
where $d_{\rm L}(z)$ is the luminosity distance to an object at redshift
$z$.
The efficiency $\epsilon (P)$ is for 64 ms trigger timescale taken from
the BATSE current burst catalogue.
In the catalogue the trigger efficiency is given only above $P\sim 0.8$
cm${}^{-2}$ s${}^{-1}$, though the data exist at lower flux range.
Hence, we restrict data fitting only to this high flux region, where the
trigger efficiency is given in the BATSE catalogue.

The relation between peak luminosity $L$ and observed peak flux $P$ (and
also with redshift $z$) is obtained if we assume the differential
rest-frame photon luminosity of the source $S(E)$ (in units of
s${}^{-1}$ keV${}^{-1}$).
As $S(E)$, we use single cutoff power-law model
\begin{equation}
 S(E)\propto E^{\beta}
  \exp\left[-\frac{(\beta +2)E}{E_{\rm peak}}\right]\,,
  \label{eq:photon spectrum}
\end{equation}
with observationally inferred value for short GRBs \cite{Ghirlanda03},
i.e., $\beta =-0.58, E_{\rm peak}=355$ keV.
The peak photon flux $P$ observed at the Earth in the energy band
$E_{\rm min}<E<E_{\rm max}$ and emitted by an isotropically radiating
source at redshift $z$ can be written by
\begin{equation}
 P=\frac{(1+z)\int_{(1+z)E_{\rm min}}^{(1+z)E_{\rm max}}S(E)\rmd E}
  {4\pi d_{\rm L}^2(z)}\,,
  \label{eq:photon flux}
\end{equation}
where we adopt $E_{\rm min}=50$ keV and $E_{\rm max}=300$ keV, which are
adjusted to the energy band of the BATSE detector.
On the other hand, an `isotropic equivalent' burst peak luminosity is
defined in the photon energy 30--2000 keV as
\begin{equation}
 L=\int_{\rm 30\,keV}^{\rm 2000\,keV}
  ES(E)\rmd E\,,
  \label{eq:photon luminosity}
\end{equation}
from which the relation connecting luminosity with peak flux and
redshift, $L(P,z)$, can be derived, combined with equations
\eref{eq:photon spectrum} and \eref{eq:photon flux}.

We simply assume that the burst luminosity distribution does not evolve
with redshift and adopt simple functional form for $\psi (L)$,
\begin{equation}
 \psi (L)=\frac{1}{L_0\Gamma(-\gamma -1)}\left(\frac{L}{L_0}\right)
  ^\gamma \exp\left(-\frac{L_0}{L}\right)\,,
  \label{eq:luminosity function}
\end{equation}
where $L$ denotes the isotropic peak luminosity in the 30--2000 keV
energy range at burst rest frame, as defined by equation \eref{eq:photon
luminosity}, $\gamma$ is the asymptotic slope at the bright end, $L_0$
marks a characteristic cutoff scale.
We label this peak LF as single power-law (SPL) model.
For comparison, we also adopt another LF whose functional form is
represented by
\begin{eqnarray}
 \psi(L)&=&\frac{1}{L_\ast}
  \left[\frac{1}{\delta_1}\left(1-\Delta_1^{-\delta_1}\right)
 +\frac{1}{\delta_2}\left(\Delta_2^{\delta_2}-1\right)\right]^{-1}
 \nonumber\\
 &&{}\times
 \left\{
  \begin{array}{ccc}
   0 & {\rm for} & L<L_\ast/\Delta_1\\
   \left(L/L_\ast\right)^{\delta_1-1} & {\rm for} 
    & L_\ast /\Delta_1<L<L_\ast\\
   \left(L/L_\ast\right)^{\delta_2-1} & {\rm for}
    & L_\ast <L<\Delta_2L_\ast\\
   0 & {\rm for} & \Delta_2L_\ast<L
    \label{eq:}
  \end{array}\right.\, ,
\end{eqnarray}
where we assume $\Delta_1=30$ and $\Delta_2=10$.
In contrast with the previous SPL-LF [equation \eref{eq:luminosity
function}], this LF is labeled as double power-law (DPL) model.
Further as two power-law indexes $\delta_1$ and $\delta_2$, we choose
two sets of their values, i.e., $(\delta_1,\delta_2) = (-0.1,-2.0),
(-0.6,-3.0)$, and name each model as DPL1 and DPL2, respectively; this
choice of the power-law index was actually adopted in the recent
publication \cite{Guetta03}, which focused on the long GRBs, on the
contrary to the present study.
For both LFs, a proper normalization $\int_0^\infty\psi (L)\rmd L=1$ is
ensured.
All the information given above allows us to calculate theoretically the
peak flux distribution using equation \eref{eq:N-P distribution}, and
to compare it with the observational data points.

In consequence, we prepare 28 models, for which the parameter fitting is
performed; each model is characterized by (1) the adopted values for
merger time distribution $\alpha$ and $\tau$, (2) the adopted LFs and
(3) cosmic SFRs.
We summarize all the models considered from this point on in
\tref{table:models}, and we label each model by the model number 1--28
as shown in the table.
\Table{Models used for parameter fittings.\label{table:models}}
\br
Model & \centre{1}{$\tau_{\rm Myr}$} & \centre{1}{$\alpha$} &
SFR & LF & Free parameters \\
\mr
\01 & 20 & $-0.5$ & SF2 & SPL & $\rho_0,\gamma,L_0$\\
\02 & 20 & $-0.7$ & SF2 & SPL & $\rho_0,\gamma,L_0$\\
\03 & 20 & $-1.0$ & SF2 & SPL & $\rho_0,\gamma,L_0$\\
\04 & 20 & $-1.5$ & SF2 & SPL & $\rho_0,\gamma,L_0$\\
\05 & \01 & $-0.5$ & SF2 & SPL & $\rho_0,\gamma,L_0$\\
\06 & \01 & $-0.7$ & SF2 & SPL & $\rho_0,\gamma,L_0$\\
\07 & \01 & $-1.0$ & SF2 & SPL & $\rho_0,\gamma,L_0$\\
\08 & \01 & $-1.5$ & SF2 & SPL & $\rho_0,\gamma,L_0$\\
\09 & \00.1 & $-0.5$ & SF2 & SPL & $\rho_0,\gamma,L_0$\\
10 & \00.1 & $-0.7$ & SF2 & SPL & $\rho_0,\gamma,L_0$\\
11 & \00.1 & $-1.0$ & SF2 & SPL & $\rho_0,\gamma,L_0$\\
12 & \00.1 & $-1.5$ & SF2 & SPL & $\rho_0,\gamma,L_0$\\
13 & \01 & $-0.5$ & SF2 & DPL1 & $\rho_0,L_\ast$\\
14 & \01 & $-0.7$ & SF2 & DPL1 & $\rho_0,L_\ast$\\
15 & \01 & $-1.0$ & SF2 & DPL1 & $\rho_0,L_\ast$\\
16 & \01 & $-1.5$ & SF2 & DPL1 & $\rho_0,L_\ast$\\
17 & \01 & $-0.5$ & SF2 & DPL2 & $\rho_0,L_\ast$\\
18 & \01 & $-0.7$ & SF2 & DPL2 & $\rho_0,L_\ast$\\
19 & \01 & $-1.0$ & SF2 & DPL2 & $\rho_0,L_\ast$\\
20 & \01 & $-1.5$ & SF2 & DPL2 & $\rho_0,L_\ast$\\
21 & \01 & $-0.5$ & SF1 & SPL & $\rho_0,\gamma,L_0$\\
22 & \01 & $-0.7$ & SF1 & SPL & $\rho_0,\gamma,L_0$\\
23 & \01 & $-1.0$ & SF1 & SPL & $\rho_0,\gamma,L_0$\\
24 & \01 & $-1.5$ & SF1 & SPL & $\rho_0,\gamma,L_0$\\
25 & \01 & $-0.5$ & SF3 & SPL & $\rho_0,\gamma,L_0$\\
26 & \01 & $-0.7$ & SF3 & SPL & $\rho_0,\gamma,L_0$\\
27 & \01 & $-1.0$ & SF3 & SPL & $\rho_0,\gamma,L_0$\\
28 & \01 & $-1.5$ & SF3 & SPL & $\rho_0,\gamma,L_0$\\
\br
\end{tabular}
\end{indented}
\end{table}

\section{Results of parameter fitting}
\label{sec:Results of parameter fitting}

\begin{figure}[htbp]
\begin{center}
\includegraphics[width=15cm]{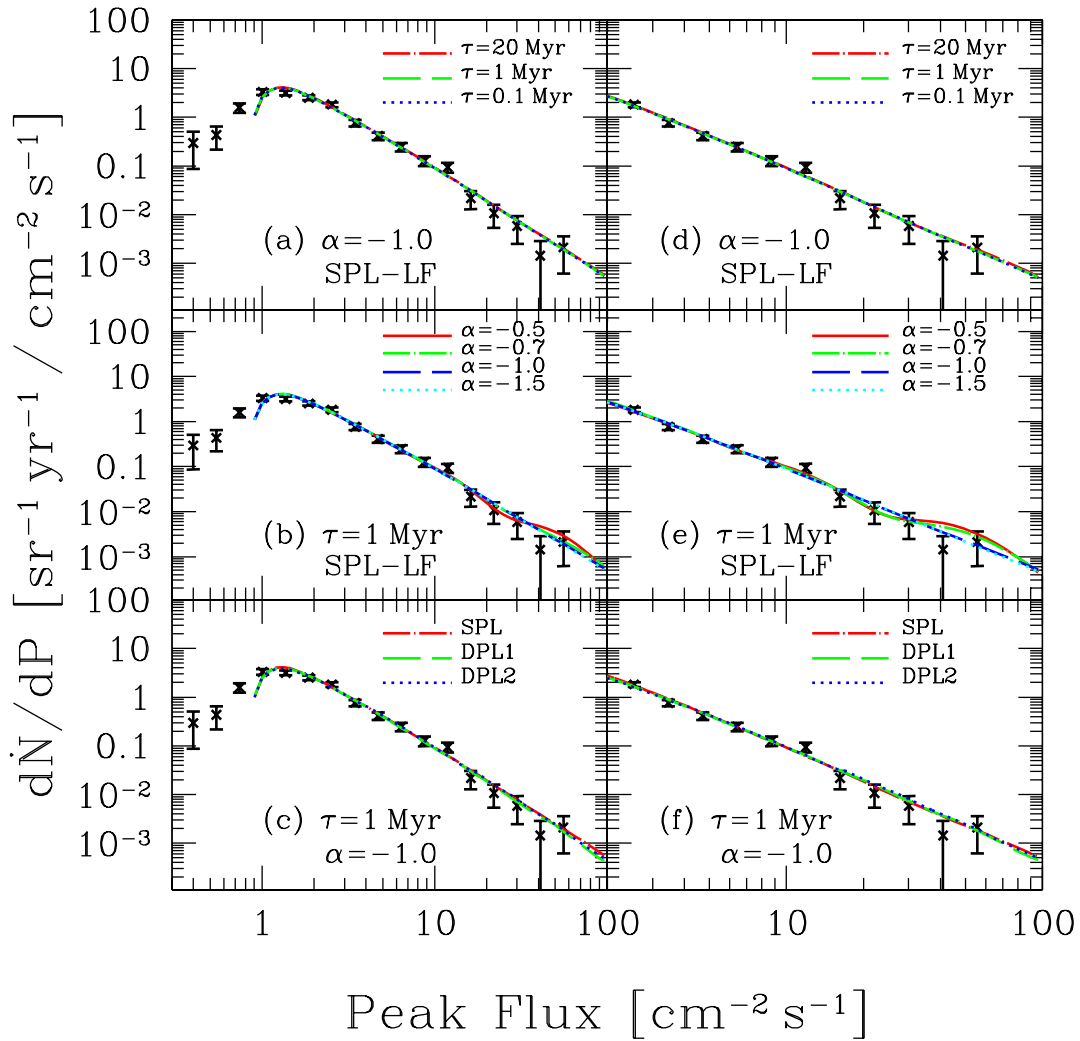}
\caption{Differential short GRB number counts as a function of peak
 photon flux. Curves represent the best fit models obtained from
 the parameter fitting to the data points. In each panel, several models
 (definitions are given in \tref{table:models}) are shown: (a) models 3,
 7 and 11, (b) models 5--8, (c) models 7, 15 and 19. Panels (d)--(f) are
 the same as (a)--(c) but the fit is restricted to the 12 data points
 from the highest flux value [compared with 15 points in the panels
 (a)--(c); see text for more details]. \label{fig:PND}}
\end{center}
\end{figure}

For each model given in \tref{table:models} we set the best fit value
of several free parameters, i.e., $\rho_0,\gamma,L_0$ in the case of
SPL-LF and $\rho_0,L_\ast$ in the case of DPL-LF, where $\rho_0$
represents the local ($z=0$) short GRB rate.
Observational data selected as short GRBs using the criterion $T_{90}<2$
s are converted into rates per unit time per unit solid angle by
considering the effective coverage of $\sim 0.48$.
We show in \fref{fig:PND} the peak flux rate distribution with the best
fit theoretical models calculated by equation \eref{eq:N-P
distribution}.
For figures \ref{fig:PND}(a)--(c), data points were used for fitting
when their flux is larger than $\sim 0.8$ cm${}^{-2}$ s${}^{-1}$, above
which the trigger efficiency is given in the BATSE current catalogue;
there, 15 data points exist and thus it results in degree of freedom
($\nu$) of 12 and 13, in the case of SPL-LF and DPL-LF, respectively.

\Fref{fig:PND}(a) shows the result of parameter fitting for the models
3, 7 and 11; i.e., we investigated the dependence on the lower cutoff
time-scale by fixing $\alpha$ to be $-1.0$.
The dependence on the power-law index of merger time distribution
$\alpha$, on the other hand, is shown in \fref{fig:PND}(b) with fixed
value of $\tau_{\rm Myr}=1$, i.e., comparison among models 5--8.
The best-fit values to three free parameters, $\rho_0,\gamma$ and
$L_0$, as well as their allowed ranges at 68\% confidence level are
summarized in \tref{table:fitting}, for models 1--12.
\Table{Best fit parameters and one sigma errors for models 1--12. They
are characterized by the different values of $(\tau,\alpha)$. As for
other properties, SF2 and SPL-LF are adopted. The values of the
minimized $\chi^2$ as well as goodness of fit are shown in the last two
columns; the degree of freedom is 12. \label{table:fitting}}
\br
& & & \centre{1}{$\rho_0$} & & \centre{1}{$L_0$}
& & \\
Model & $\tau_{\rm Myr}$ & \centre{1}{$\alpha$} & (yr${}^{-1}$
Gpc${}^{-3}$) & \centre{1}{$\gamma$} & ($10^{51}$ erg s${}^{-1}$) &
\centre{1}{$\chi^2/\nu$} & \centre{1}{g.o.f.}\\
\mr
\01 & 20 & $-0.5$ & $1.3\0\pm 0.3$ & $-6.4^{+2.4}_{-3.3}$ &
$4.6^{+3.1}_{-2.5}$ & 16.8/12 & 0.16 \\
\02 & 20 & $-0.7$ & $0.81^{+0.24}_{-0.16}$ & $-4.4^{+1.3}_{-1.9}$ &
$3.2^{+3.1}_{-2.0}$ & 17.1/12 & 0.15 \\
\03 & 20 & $-1.0$ & $0.41^{+0.14}_{-0.15}$ & $-3.2^{+0.6}_{-1.8}$ &
$2.3^{+5.5}_{-1.9}$ & 17.2/12 & 0.14 \\
\04 & 20 & $-1.5$ & $0.18^{+0.39}_{-0.07}$ & $-2.8^{+0.6}_{-0.9}$ &
$2.1^{+4.9}_{-1.9}$ & 17.3/12 & 0.14 \\
\05 & \01 & $-0.5$ & $1.2\pm 0.3$ & $-6.2^{+2.3}_{-3.3}$ &
$4.5^{+3.1}_{-2.4}$ & 16.8/12 & 0.16 \\
\06 & \01 & $-0.7$ & $0.72^{+0.23}_{-0.14}$ & $-4.0^{+1.1}_{-1.6}$ &
$3.0^{+3.5}_{-3.0}$ & 17.1/12 & 0.14 \\
\07 & \01 & $-1.0$ & $0.31\pm 0.12$ & $-3.0^{+0.5}_{-1.6}$ &
$2.2^{+6.1}_{-2.0}$ & 17.3/12 & 0.14 \\
\08 & \01 & $-1.5$ & $0.14^{+0.27}_{-0.06}$ & $-2.7^{+0.5}_{-1.0}$ &
$2.2^{+4.9}_{-1.9}$ & 17.3/12 & 0.14 \\
\09 & \00.1 & $-0.5$ & $1.2\0\pm 0.3$ & $-6.2^{+2.3}_{-3.3}$ &
$4.5^{+3.1}_{-2.5}$ & 16.8/12 & 0.16 \\
10 & \00.1 & $-0.7$ & $0.70^{+0.22}_{-0.14}$ & $-4.0^{+1.1}_{-1.6}$ &
$3.0^{+3.5}_{-2.6}$ & 17.1/12 & 0.14 \\
11 & \00.1 & $-1.0$ & $0.27\pm 0.11$ & $-2.9^{+0.5}_{-1.5}$ &
$2.2^{+5.5}_{-2.0}$ & 17.3/12 & 0.14 \\
12 & \00.1 & $-1.5$ & $0.13^{+0.25}_{-0.06}$ & $-2.7^{+0.5}_{-1.0}$ &
$2.2^{+5.1}_{-1.9}$ & 17.3/12 & 0.14 \\
\br
\end{tabular}
\end{indented}
\end{table}
In the same table, we also show the values of $\chi^2$ as well as a
goodness of fit (g.o.f.) in the last two columns.
The value of $\chi^2$ is not sufficiently small compared to the degree
of freedom, but the fit is still acceptable.
As clearly seen from \tref{table:fitting}, the dependence on the lower
cutoff time $\tau$ is very weak, while the values of the best-fit
parameters are rather sensitve to the power-law index $\alpha$.
These characteristics obviously reflects the strong (weak) dependence of
the short GRB rate on $\alpha$ ($\tau$), which is already discussed in
\sref{sec:Formation history of short gamma-ray bursts} and shown in
figures \ref{fig:GRBrate1} and \ref{fig:GRBrate2}.
\Tref{table:fitting} suggests that at present, with any models we can
obtain reasonable fit to the data; no models are highly favoured or
disfavoured compared to the others.

Several implications for the burst parameters, i.e., $\rho_0,\gamma$
and $L_0$, are also obtained by the fit.
The local short GRB rate is found to be 0.13--1.3 yr${}^{-1}$
Gpc${}^{-3}$, which is comparable to or in most cases larger than the
local long GRB rate inferred by the analysis in reference
\cite{Porciani01}, i.e., 0.12 yr${}^{-1}$ Gpc${}^{-3}$, by some factor
to one order of magnitude.
Although only 1/3 fraction of total number of detected GRBs is short
type, the current estimated formation rate of short GRBs dominates over
that of long GRBs.
The remaining parameters related to the LF can compensate this apparent
conflict.
The best fit values for such parameters are $-6.4<\gamma<-2.7$, and
(2.1--4.6)$\times 10^{51}$ erg s${}^{-1}$ for $L_0$.
According to reference \cite{Porciani01}, on the other hand for long
GRBs, the best fit values for these parameters are $L_0 = 6\times
10^{51}$ erg s${}^{-1}$ and $\gamma = -2.9$, with which the luminosity
distribution of long GRBs is expected to be brighter and flatter (owing
to larger $L_0$ and $\gamma$) than that of short GRBs.
Thus, smaller values for $L_0$ and $\gamma$, obtained for short GRBs,
are expected to push more GRBs below the detection threshold, reducing
the total number of detected GRB rate.
Therefore, the resulting number of short GRBs is smaller than that of
long ones, although the local rate itself $\rho_0$ is dominated by the
short population.

We further show the same results but for models 13--20 and 21--28 in
tables \ref{table:fitting LF} and \ref{table:fitting SFR}, respectively.
\Table{Best fit parameters and one sigma errors for models 13--20. They
are characterized by the adopted LFs and different values of
$\alpha$. As for other properties, SF2 and $\tau_{\rm Myr}=1$ are
adopted. The values of the minimized $\chi^2$ as well as goodness of fit
are shown in the last two columns; the degree of freedom is
13. \label{table:fitting LF}}
\br
& & & \centre{1}{$\rho_0$} & \centre{1}{$L_\ast$} & & \\
Model & LF & \centre{1}{$\alpha$} & (yr${}^{-1}$ Gpc${}^{-3}$) &
($10^{51}$ erg s${}^{-1}$) & \centre{1}{$\chi^2/\nu$} &
\centre{1}{g.o.f.}\\
\mr
13 & DPL1 & $-0.5$ & $1.8\0^{+0.5}_{-0.1}$ & $\02.0^{+0.2}_{-1.3}$ &
18.3/13 & 0.15 \\
14 & DPL1 & $-0.7$ & $1.2^{+0.2}_{-0.1}$ & $\02.1\pm 0.3$ & 16.4/13 &
0.23 \\
15 & DPL1 & $-1.0$ & $0.40^{+0.06}_{-0.07}$ & $\03.9^{+0.9}_{-0.5}$ &
16.5/13 & 0.22 \\
16 & DPL1 & $-1.5$ & $0.15^{+0.04}_{-0.02}$ & $\06.2^{+0.7}_{-1.6}$ &
17.4/13 & 0.18 \\
17 & DPL2 & $-0.5$ & $2.0\0^{+0.5}_{-0.3}$ & $\03.7\pm 0.6$ & 16.2/13 &
0.24 \\
18 & DPL2 & $-0.7$ & $1.3\0^{+0.1}_{-0.2}$ & $\04.3^{+0.8}_{-0.9}$ &
14.8/13 & 0.32 \\
19 & DPL2 & $-1.0$ & $0.36^{+0.07}_{-0.04}$ & $\09.3^{+0.6}_{-1.4}$ &
16.4/13 & 0.23 \\
20 & DPL2 & $-1.5$ & $0.15\pm 0.01$ & $13.4^{+0.3}_{-0.1}$ & 15.5/13 &
0.28 \\
\br
\end{tabular}
\end{indented}
\end{table}
\Table{Best fit parameters and one sigma errors for models 21--28. They
are characterized by the adopted SFRs and different values of
$\alpha$. As for other properties, SPL-LF and $\tau_{\rm Myr}=1$ are
adopted. The values of the minimized $\chi^2$ as well as goodness of fit
are shown in the last two columns; the degree of freedom is
12. \label{table:fitting SFR}}
\br
& & & \centre{1}{$\rho_0$} & & \centre{1}{$L_0$} & & \\
Model & SFR & \centre{1}{$\alpha$} & (yr${}^{-1}$ Gpc${}^{-3}$) &
\centre{1}{$\gamma$}& ($10^{51}$ erg s${}^{-1}$) &
\centre{1}{$\chi^2/\nu$} & \centre{1}{g.o.f.}\\
\mr
21 & SF1 & $-0.5$ & $1.2\0^{+0.3}_{-0.2}$ & $-6.2^{+2.2}_{-3.3}$ &
$4.5^{+3.0}_{-2.4}$ & 17.0/12 & 0.16 \\
22 & SF1 & $-0.7$ & $0.72^{+0.21}_{-0.13}$ & $-3.9^{+1.1}_{-1.2}$ &
$2.8^{+3.6}_{-2.3}$ & 17.3/12 & 0.14 \\
23 & SF1 & $-1.0$ & $0.32^{+1.14}_{-0.11}$ & $-2.9^{+0.8}_{-0.9}$ &
$1.9^{+3.0}_{-1.8}$ & 17.4/12 & 0.13 \\
24 & SF1 & $-1.5$ & $0.15^{+0.37}_{-0.06}$ & $-2.6^{+0.5}_{-0.6}$ &
$1.7^{+2.7}_{-1.6}$ & 17.4/12 & 0.13 \\
25 & SF3 & $-0.5$ & $1.3\0\pm 0.3$ & $-6.4^{+2.4}_{-3.3}$ &
$4.6^{+3.1}_{-2.4}$ & 16.7/12 & 0.16 \\
26 & SF3 & $-0.7$ & $0.77^{+0.26}_{-0.16}$ & $-4.2^{+1.2}_{-1.9}$ &
$3.3^{+2.9}_{-2.1}$ & 17.0/12 & 0.15 \\
27 & SF3 & $-1.0$ & $0.32^{+0.13}_{-0.16}$ & $-3.1^{+0.6}_{-4.9}$ &
$2.5^{+18.8}_{-2.5}$ & 17.2/12 & 0.14 \\
28 & SF3 & $-1.5$ & $0.16^{+0.28}_{-0.17}$ & $-2.8^{+0.6}_{-1.9}$ &
$2.7^{+10.5}_{-2.3}$ & 17.3/12 & 0.14 \\
\br
\end{tabular}
\end{indented}
\end{table}
The best-fit flux distribution for the DPL-LF models (model 15 and 19)
are compared with that for the SPL-LF model 7 in \fref{fig:PND}(c).
In the case of the DPL-LF models, the quality of the fit becomes
slightly better, but not to the extent that we can conclude that the
short GRB data strongly favours DPL luminosity distribution.
The adopted SFR does not give any significant difference from one
another; this clearly indicates that at present the local contribution
$z<1.5$ from SFRs dominates the bulk of data.
Because many observations with various wavebands give roughly consistent
results with the SF1, SF2 and SF3 models at low-redshift region, i.e.,
the SFR increases from the current epoch $z=0$ at least to $z=1$, we
believe that the uncertainties concerning the adopted SFR are small at
low redshift and therefore giving no significant ambiguity to the
parameter values as found from the comparison between tables
\ref{table:fitting} and \ref{table:fitting SFR}.

Until this point, we used all the available data given in the BATSE
current burst catalogue.
However, it should be noted that because the trigger efficiency used in
the present paper does not include the correction concerning the effects
of atmospheric scattering.
As the result, the efficiencies at low fluxes are underestimated as also
mentioned in the BATSE catalogue itself, although we do not know whether
our calculations given above are affected by such an effect.
In order to investigate that point, we repeated the same procedure given
above, but with restricting the fit to the 12 data points from the high
values of peak flux (compared with the 15 data points previously
adopted), and its result is summarized in figures \ref{fig:PND}(d)--(f),
tables \ref{table:fitting_large} and \ref{table:fitting large LF}.
\Table{The same as \tref{table:fitting}, but the fitting is restricted
to the rather high-flux region. The degree of freedom is reduced to
9. \label{table:fitting_large}}
\br
& & & \centre{1}{$\rho_0$} & & \centre{1}{$L_0$} & & \\
Model & $\tau_{\rm Myr}$ & \centre{1}{$\alpha$} & (yr${}^{-1}$
Gpc${}^{-3}$) & \centre{1}{$\gamma$} & ($10^{51}$ erg s${}^{-1}$) &
\centre{1}{$\chi^2/\nu$} & \centre{1}{g.o.f.}\\
\mr
\01 & 20 & $-0.5$ & $1.5\0\pm 0.4$ & $-7.2^{+2.7}_{-3.2}$ &
$5.0^{+3.2}_{-2.4}$ & 8.9/9 & 0.44 \\
\02 & 20 & $-0.7$ & $1.2\0\pm 0.4$ & $-6.8^{+2.7}_{-3.3}$ &
$4.8^{+3.1}_{-2.7}$ & 8.5/9 & 0.49 \\
\03 & 20 & $-1.0$ & $0.51^{+0.36}_{-0.19}$ & $-3.8^{+0.6}_{-4.3}$ &
$2.9^{+3.1}_{-2.1}$ & 8.9/9 & 0.45 \\
\04 & 20 & $-1.5$ & $0.21^{+0.23}_{-0.15}$ & $-3.0^{+0.6}_{-40.2}$ &
$2.3^{+173.2}_{-2.0}$ & 9.0/9 & 0.44 \\
\05 & \01 & $-0.5$ & $1.4\0\pm 0.4$ & $-7.2^{+2.7}_{-3.2}$ &
$5.0^{+3.1}_{-2.5}$ & 8.9/9 & 0.45 \\
\06 & \01 & $-0.7$ & $1.1\0\pm 0.4$ & $-6.6^{+2.7}_{-3.4}$ &
$4.8^{+3.1}_{-2.8}$ & 8.4/9 & 0.49 \\
\07 & \01 & $-1.0$ & $0.38^{+0.33}_{-0.27}$ & $-3.4^{+0.8}_{-48.4}$ &
$2.6^{+210.4}_{-2.6}$ & 9.0/9 & 0.44 \\
\08 & \01 & $-1.5$ & $0.17^{+0.21}_{-0.13}$ & $-2.9^{+0.5}_{-6.4}$ &
$2.3^{+41.8}_{-2.0}$ & 9.0/9 & 0.43 \\
\09 & \00.1 & $-0.5$ & $1.4\0\pm 0.4$ & $-7.2^{+2.7}_{-3.2}$ &
$5.0^{+3.1}_{-2.5}$ & 8.9/9 & 0.45 \\
10 & \00.1 & $-0.7$ & $1.1\0\pm 0.4$ & $-6.6^{+2.8}_{-3.5}$ &
$4.7^{+3.1}_{-2.8}$ & 8.5/9 & 0.49 \\
11 & \00.1 & $-1.0$ & $0.33^{+0.32}_{-0.24}$ & $-3.2^{+0.7}_{-47.9}$ &
$2.5^{+207.5}_{-2.1}$ & 9.0/9 & 0.43 \\
12 & \00.1 & $-1.5$ & $0.16^{+0.19}_{-0.12}$ & $-2.9^{+0.5}_{-4.8}$ &
$2.3^{+40.8}_{-2.0}$ & 9.0/9 & 0.43 \\
\br
\end{tabular}
\end{indented}
\end{table}
\Table{The same as \tref{table:fitting LF}, but the fitting is
restricted to the rather high-flux region. The degree of freedom is
reduced to 10. \label{table:fitting large LF}}
\br
& & & \centre{1}{$\rho_0$} & \centre{1}{$L_\ast$} & & \\
Model & LF & \centre{1}{$\alpha$} & (yr${}^{-1}$ Gpc${}^{-3}$) &
($10^{51}$ erg s${}^{-1}$) & \centre{1}{$\chi^2/\nu$} &
\centre{1}{g.o.f.}\\
\mr
13 & DPL1 & $-0.5$ & $1.8\0^{+0.3}_{-0.2}$ & $\02.0^{+0.3}_{-1.6}$ &
10.9/10 & 0.37 \\
14 & DPL1 & $-0.7$ & $1.3\0\pm 0.2$ & $\02.1\pm 0.3$ & \08.9/10 & 0.54
\\
15 & DPL1 & $-1.0$ & $0.42^{+0.40}_{-0.10}$ & $\03.9^{+1.0}_{-2.1}$ &
\08.7/10 & 0.56 \\
16 & DPL1 & $-1.5$ & $0.15^{+0.15}_{-0.04}$ & $\06.2^{+1.0}_{-2.7}$ &
\08.9/10 & 0.54 \\
17 & DPL2 & $-0.5$ & $2.0\0^{+1.1}_{-0.4}$ & $\03.7^{+0.8}_{-1.9}$ &
\09.8/10 & 0.46 \\
18 & DPL2 & $-0.7$ & $1.6\0^{+0.8}_{-0.4}$ & $\03.7^{+1.3}_{-1.7}$ &
\08.3/10 & 0.60 \\
19 & DPL2 & $-1.0$ & $0.37^{+0.09}_{-0.12}$ & $\09.3^{+3.0}_{-1.4}$ &
\09.6/10 & 0.48 \\
20 & DPL2 & $-1.5$ & $0.15^{+0.08}_{-0.02}$ & $13.0^{+0.5}_{-4.0}$ &
\08.6/10 & 0.57 \\
\br
\end{tabular}
\end{indented}
\end{table}
This time, the degree of freedom is reduced to 9 for SPL-LF and 10 for
DPL-LFs.
It is noticeable that the quality of the fit becomes considerably better
for both SPL and DPL models.
We believe that this fact suggests that the fitting improvement is not
due to the wrong modeling (at least of LFs) but rather due to exclusion
of lower-flux data points, which may be affected by the atmospheric
scatterings.
Although the values of the best-fit parameters are quite consistent with
the previously obtained values in tables \ref{table:fitting} and
\ref{table:fitting LF}, the accompanying errors become significantly
large.
Further, every model gives reasonable fit at almost the same level; no
models can be excluded or restricted by the present analysis.

\section{Discussion}
\label{sec:Discussion}

\subsection{Implications for jet structure and intrinsic luminosity}
\label{sub:Implications for jet structure and intrinsic luminosity}

It is of interest to compare these properties with those inferred from
other theoretical/observational approaches.
The local merger rate of binary NS-NS or NS-BH sysyem calculated by
population synthesis code ranges by many orders of magnitude.
For instance according to reference \cite{Fryer99b}, it ranges 0.01--80
Myr${}^{-1}$ galaxy${}^{-1}$ depending on the various parameters, with
1 Myr${}^{-1}$ galaxy${}^{-1}$ in the case of standard choice of
parameter set.
On the other hand, more recent study \cite{Belczynski02b} claims that
larger number of NS-NS merger can be created; the coalescence rate
ranges 1--300 Myr${}^{-1}$ galaxy${}^{-1}$ with standard value of 50
Myr${}^{-1}$ galaxy${}^{-1}$.

In this section, we compare these merger rates obtained from their
standard models, i.e., 1 Myr${}^{-1}$ galaxy${}^{-1}$ \cite{Fryer99b} or
50 Myr${}^{-1}$ galaxy${}^{-1}$ \cite{Belczynski02b} with our
inference of $\rho_0$ using short GRB observations.
We simply convert the rate per galaxy into that per volume assuming that
the number density of galaxies is $\sim 10^{-2}$ Mpc${}^{-3}$.
Using this conversion factor, the rate 1 (50) Myr${}^{-1}$
galaxy${}^{-1}$ corresponds to 10 (500) yr${}^{-1}$ Gpc${}^{-3}$, with
which the values $\rho_0$ we have obtained in the previous section
should be compared.
From tables \ref{table:fitting}--\ref{table:fitting large LF}, we take
$\rho_0=0.5$ yr${}^{-1}$ Gpc${}^{-3}$ as our reference value, which is
significantly smaller than those derived from the population synthesis
approach; our evaluated value is 0.05 and $10^{-3}$ of the local merger
rate from references \cite{Fryer99b} and \cite{Belczynski02b},
respectively.
However, this disagreement may be a natural consequence of the jet
structure of GRBs or of the fact that not every merger produces GRBs.
Assuming that all the binary mergers containing at least one neutron
star certainly produce short GRBs, we can suggest the beaming angle of
the jet using the rate difference obtained above.
Considering that the fraction of apparent GRB rate is only $\theta^2/2$
of the intrinsic GRB rate (coalescence rate), the derived values for the
jet opening angle $\theta$ are 18${}^\circ$ and 2.6${}^\circ$, which are
derived from the ratio of $\rho_0$ obtained from our approach to that
from the local coalescence rate taken from references \cite{Fryer99b}
and \cite{Belczynski02b}, respectively.
These inferences of the jet opening angles are quite well consistent
with those for long GRBs determined by the achromatic break of the
afterglow light curves \cite{Frail01}.
The intrinsic values corresponding to the luminosities, $L_0$ and
$L_\ast$, are suggested to be also smaller than the values summarized in
tables \ref{table:fitting}--\ref{table:fitting large LF} by a factor of
$0.05$ or $10^{-3}$, depending on whether the opening angle is
18${}^\circ$ or 2.6${}^\circ$.
We summarize these particulars in \tref{table:suggestion}, but in which
the specific model 7, i.e., $(\tau_{\rm Myr},\alpha)=(1,-1.0)$, SF2 and
SPL-LF, is used for the evaluation with the values given in
\tref{table:fitting}.
\Table{Comparison with other theoretical/observational rate
estimates. The galactic merger rate is converted into global rate with
an assumption that the local galaxy density is $10^{-2}$
Mpc${}^{-3}$. As the local short GRB rate $\rho_0$ and cutoff luminosity
$L_0$, the values given in \tref{table:fitting} are adopted with the
specific model 7. \label{table:suggestion}}
\br
 & & \centre{1}{Merger rate} & \centre{1}{Jet angle} &
 \centre{1}{$L_0^{\rm int}$} \\
 \centre{1}{Method} & \centre{1}{Reference} & \centre{1}{(Myr${}^{-1}$
 Gal${}^{-1}$)} & \centre{1}{$\theta~({}^\circ)$} &
 \centre{1}{($10^{49}$ erg s${}^{-1}$)} \\
\mr
Population synthesis & Fryer \etal \cite{Fryer99b} & \0\01.2 & 13 & 5.7
\\
 & Belczynski \etal \cite{Belczynski02b}& \052.7 & \02.0 & 0.13 \\
Observation & Kalogera \etal \cite{Kalogera03} & 180 & \01.1 & 0.038
\\
\br
\end{tabular}
\end{indented}
\end{table}
In the same table, we also show the values obtained by comparing our
$\rho_0$ with the observationally inferred local coalescence rate, which
was recently found to be rather large as $180^{+477}_{-144}$
Myr${}^{-1}$ galaxy${}^{-1}$, although the estimate contains a fair
amount of uncertainty \cite{Kalogera03}.
In this case, the obtained jet angle is a little bit small but is still
marginally consistent with that of long GRBs observed by the achromatic
break of afterglow light curves.
In consequence, although there is no direct evidence supporting the
connection between short GRBs and binary neutron star mergers, we can
explain observed data quite naturally with such a connection as well as
jet-like nature of short GRBs.

In the above discussion, we have assumed that the jet structure is {\it
uniform} in the opening angle, i.e., the kinetic energy does not depend
on the solid angle.
On the other hand, several models that assume {\it structured} jet,
i.e., the kinetic energy depends on the solid angle, are proposed (e.g.,
see references \cite{Rossi02,Zhang02}), and whether the jet is uniform
or structured is matter of controversy \cite{Guetta03,Perna03}.
At the end of this subsection, therefore, we consider the possible case
of structured jet model; we adopt the power-law distribution of kinetic
energy per unit solid angle, i.e., $\epsilon(\theta)\propto\theta^
{-k}$ and $k=2$ as the canonical value \cite{Rossi02}.
A remarkable property of the structured jet model is that the isotropic
LF is determined to be $\psi (L)\propto L^{-1-2/k}$, because the
apparent luminosity depends on the direction alone
\cite{Rossi02,Zhang02}.
In our calculation, this corresponds to the case of SPL-LF model
summarized in \tref{table:fitting}.
From the table, the canonical value of $\gamma=-2$, expected from the
structured jet model with $k=2$, is not favoured, while the statistical
significance is not large.
Steeper jet structure (due to larger $k$) gives flatter LF (larger
$\gamma$) and it is further more disfavoured.
In addition, our results are also against the quasi-universal Gaussian
jet structured model, which predicts $\psi(L)\propto L^{-1}$ even if
some dispersion of parameter values is allowed
\cite{Zhang02,Lloyd-Ronning04}.

\subsection{Inferred redshift distribution and intrinsic burst duration}
\label{sub:Inferred redshift distribution and intrinsic burst duration}

\begin{figure}[htbp]
\begin{center}
\includegraphics[width=10cm]{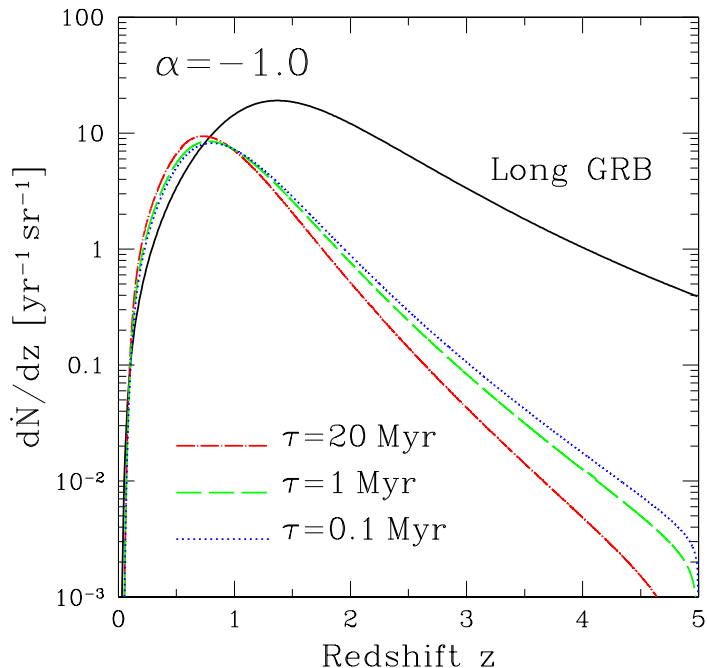}
\caption{Redshift distribution of short GRBs detected by BATSE. Three
 models 3, 7 and 11, which are different by the value of lower cutoff
 time-scale $\tau$ with $\alpha$ fixed to be $-1.0$, are compared with
 each other. The distribution of long GRBs obtained by parameters given
 in reference \cite{Porciani01} is also shown for
 comparison. \label{fig:rho_z1}}
\end{center}
\end{figure}

\begin{figure}[htbp]
\begin{center}
\includegraphics[width=10cm]{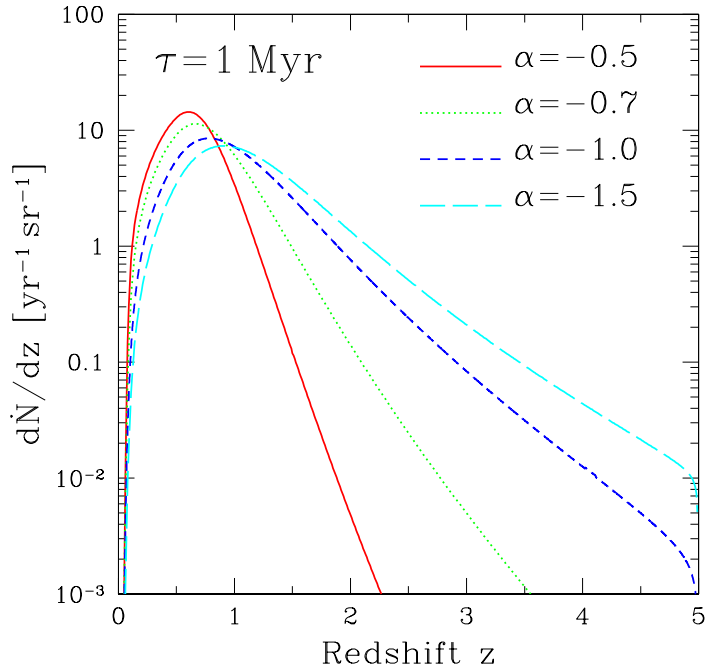}
\caption{The same as \fref{fig:rho_z1}, but evaluated for models 5--8,
 i.e., with various values of $\alpha$ with fixed $\tau$ to be 1
 Myr. \label{fig:rho_z2}}
\end{center}
\end{figure}

Although we do not have any redshift information for short GRBs, we can
infer their redshift distribution detected by BATSE, by using the
formulation,
\begin{equation}
 \frac{\rmd\dot N}{\rmd z}=\frac{\rmd V(z)}{\rmd z}
  \frac{R_{\rm GRB}(z)}{1+z}
  \int_{P_{\rm min}}^{P_{\rm max}}
  \rmd P\frac{\partial L(P,z)}{\partial P}
  \psi(L(P,z))\epsilon (P)\,.
  \label{eq:redshift distribution}
\end{equation}
Figures \ref{fig:rho_z1} and \ref{fig:rho_z2} show the redshift
distribution $\rmd\dot N/\rmd z$ calculated using equation
\eref{eq:redshift distribution} with the GRB formation rate $R_{\rm
GRB}(z)$ shown in figures \ref{fig:GRBrate1} and \ref{fig:GRBrate2}.
In \fref{fig:rho_z1}, we also show the expected redshift distribution of
long GRBs detected by BATSE using the best fit parameters taken from
reference \cite{Porciani01}, for comparison.
From the figure, the short GRBs are considered to be more localized
relative to the long ones, but its degree is relaxed when the slope of
the merger time distribution becomes steeper (for smaller $\alpha$).
These redshift distributions would be another probe of its origin if the
accompanied afterglow data were accumulated.

Although we classified the bursts with $T_{90}<2$ s as short GRBs, the
observed duration is affected by cosmological time dilation.
In order to obtain intrinsic duration distribution for short GRBs, we
use an assumption that it obeys log-normal distribution as well as the
redshift distribution of the short GRBs already obtained via equation
\eref{eq:redshift distribution}.
The rate of detected bursts per unit redshift range $\rmd z$ per unit
intrinsic duration $\rmd \Delta t_{\rm int}$ is obtained by
\begin{equation}
 \rmd^2\dot N=P(\Delta t_{\rm int})\rmd\Delta t_{\rm int}
  \frac{\rmd\dot N}{\rmd z}\rmd z\,,
\end{equation}
where $P(\Delta t_{\rm int})$ represents the probability distribution
function of intrinsic duration, which we assume log-normal type;
$\rmd\dot N/\rmd z$ has already been obtained in the previous
paragraph.
This relation can be rewritten as a function of observed duration
$\Delta t=(1+z)\Delta t_{\rm int}$ and redshift $z$, and then be
integrated over redshift as:
\begin{eqnarray}
 \frac{\rmd\dot N}{\rmd\Delta t}&=&
  \int_0^{\infty}P(\Delta t_{\rm int})\frac{\rmd\dot N}{\rmd z}
  \left.\frac{\partial\Delta t_{\rm int}}{\partial\Delta t}\right|_z
  \rmd z\nonumber\\
 &=&\int_0^{\infty}\frac{P((1+z)^{-1}\Delta t)}{1+z}
  \frac{\rmd\dot N}{\rmd z}\rmd z\,,
  \label{eq:intrinsic duration distribution}
\end{eqnarray}
with which we can obtain the values for central duration as well as the
standard deviation of the log-normal distribution $P(\Delta t_{\rm
int})$, by fitting the observed duration distribution by BATSE.
As the result we found that $\langle\Delta t_{\rm int}\rangle =
0.24$--0.31 s and $\sigma_{\log\Delta t_{\rm int}} = 0.50$--0.55,
depending on the adopted parameters of $(\tau,\alpha)$.

\section{Conclusions}
\label{sec:Conclusions}

Recent progresses concerning GRBs such as evidence for collimated jet
structure or association with supernova explosions are all restricted to
long duration GRBs with $T_{90}>2$ s.
For the short GRBs, on the other hand, because of the lack of any
afterglow information, we only poorly understand them.
However, a clear bimodal duration distribution and several numerical
simulations suggest that the origin of short GRBs is different from that
of long ones, and the binary neutron star merger event remains as a
viable candidate.

Based on the assumption that all the short GRBs with $T_{90}<2$ s are
connected to the binary neutron star mergers, we analyzed the current
available data taken from the BATSE catalogue.
In the analyses, we parameterized the merger time distribution, which is
essential for short GRB rate estimate, as $P_{\rm m}(t)\propto t^\alpha$
with lower cutoff time-scale $\tau$.
The short GRB rate, calculated using such a parameterization with
equation \eref{eq:short GRB rate} and generated from the model for
cosmic star formation rate SF2 in reference \cite{Porciani01}, was found
to be quite insensitive to the value of $\tau$, but significantly
sensitive to $\alpha$, as shown in figures \ref{fig:GRBrate1} and
\ref{fig:GRBrate2}.
With such formation rate models, the peak flux distribution observed by
BATSE was fitted.
We further adopted two different functional form for isotropic
luminosity distribution, i.e., SPL and DPL.
In the fitting process, we used several free parameters, $\rho_0,
\gamma, L_0$ ($\rho_0, L_\ast$) in the case of SPL-LF (DPL-LF) model,
where $\rho_0$ represents the local short GRB rate density and the
others are the parameters related to the LFs.
The result of parameter fitting is summarized in figures \ref{fig:PND}
as well as tables \ref{table:fitting}--\ref{table:fitting SFR}.
We found that every model gives acceptable fit with reasonable parameter
values, and that they are essentially insensitive to the adopted SFR
models (SF1--SF3) because the bulk of the data points are expected to be
attributed to local redshift region $z<1.5$.
The resulting values for the local GRB rate are found to be 0.1--2
yr$^{-1}$ Gpc$^{-3}$.
Furthermore, the quality of the fit considerably improves when we omit
several data points at low peak flux as shown in tables
\ref{table:fitting_large} and \ref{table:fitting large LF}.
This may be attributed to the fact that the BATSE triggering efficiency
is underestimated at such a region, because it does not contain the
correction to the atmospheric scatterings.

We also discussed several implications for the intrinsic properties of
possible sources of short GRBs.
(1) By comparing with the galactic coalescence rate of binary NS-NS
systems obtained by theoretical population synthesis or observations of
binary neutron star systems, and also by assuming that all such mergers
certainly produce short GRBs, we constrained the jet structure as well
as the intrinsic luminosity function.
The obtained values for a specific model is summarized in
\tref{table:suggestion}.
(2) The structured jet model, where the kinetic energy depends on the
direction, e.g., $\epsilon (\theta)\propto\theta^{-k}$, was also
considered; a remarkable characteristic of this model is that the
isotropic LF becomes SPL function $\psi (L)\propto L^{-1-2/k}$.
According to our analyses, the model with $k\ge 2$ is disfavoured
although the statistical significance is still small.
(3) Although the redshift distribution of short GRBs is not known yet
since there is no discovery of afterglow signals, we can infer it using
equation \eref{eq:redshift distribution} and best fit parameters, as
shown in figures \ref{fig:rho_z1} and \ref{fig:rho_z2}.
With this distribution and the observed duration distribution of short
GRBs, we derived the parameters for intrinsic duration distribution
which we assumed to have a log-normal property, by correcting the effect
of cosmological time dilation.

\ack
This work was supported by Grant-in-Aid for JSPS Fellows.

\section*{References}
\bibliography{refs}

\end{document}